\documentclass[a4paper,aps,prd,onecolumn,nofootinbib,preprintnumbers,superscriptaddress,amsmath,amssymb]{revtex4}

\usepackage{graphicx}
\usepackage{bm,natbib,url,textcase}
\usepackage[english]{babel}
\usepackage[T1]{fontenc}
\usepackage{braket}
\usepackage{quoting}
\quotingsetup{font=small}
\usepackage{csquotes}
\usepackage{dcolumn}
\usepackage{microtype}
\numberwithin{equation}{section}
\usepackage[colorlinks]{hyperref}
\usepackage[dvipsnames,svgnames,x11names]{xcolor}

\usepackage{multirow}
\usepackage{times}

\hypersetup{
	colorlinks=true,                
	linkcolor=red!60!black,         
	citecolor=green!55!black,       
	urlcolor=blue!40!black,         
}

\makeatletter
\renewcommand\@makefnmark{\hbox{\@textsuperscript{\normalfont\color{blue!40!black}\@thefnmark}}}
\makeatother

\usepackage{latexsym}
\def\be{\begin{equation}}
	\def\ee{\end{equation}}
\def\bea{\begin{eqnarray}}
	\def\eea{\end{eqnarray}}
\def\ben{\begin{enumerate}}
	\def\een{\end{enumerate}}

\usepackage{mathtools}
\DeclarePairedDelimiter\abs{\lvert}{\rvert}%
\makeatletter
\let\oldabs\abs
\def\abs{\@ifstar{\oldabs}{\oldabs*}}

\def\av{``}
\def\cv{''}
\def\z{~}

\def\f{\varphi}

\usepackage{tikz}
\definecolor{purple}{rgb}{1,0,1}
\definecolor{lime}{HTML}{A6CE39} 

\newcommand{\orcidicon}{%
	\begin{tikzpicture}
		\draw[lime, fill=lime] (0,0) 
		circle [radius=0.15] 
		node[white] {{\fontfamily{qag}\selectfont \tiny ID}};
		\draw[white, fill=white] (-0.0625,0.095) 
		circle [radius=0.007];
	\end{tikzpicture}	\hspace{-2mm}
}
\newcommand\orcidMarcello{{\href{https://orcid.org/0000-0003-0397-2705}{\orcidicon}}}
\newcommand\orcidDaniele{{\href{https://orcid.org/0000-0003-4379-2549}{\orcidicon}}}
\newcommand\orcidSalvatore{{\href{https://orcid.org/0000-0003-4886-2024}{\orcidicon}}}
\newcommand\orcidFrancisco{{\href{https://orcid.org/0000-0002-9388-8373}{\orcidicon}}}

\begin{document}
    
   \title{Bouncing Cosmology in Fourth-Order Gravity}
	
	\author{Marcello Miranda\orcidMarcello\!}
	\email{marcello.miranda@unina.it}
	\affiliation{Scuola Superiore Meridionale, Largo San Marcellino 10, I-80138, Napoli, Italy.}
	\affiliation{INFN Sez. di Napoli, Compl. Univ. di Monte S. Angelo, Edificio G, Via	Cinthia, I-80126, Napoli, Italy.}
	
	
	\author{Daniele Vernieri\orcidDaniele\!}
	\email{daniele.vernieri@unina.it}
	\affiliation{INFN Sez. di Napoli, Compl. Univ. di Monte S. Angelo, Edificio G, Via	Cinthia, I-80126, Napoli, Italy.}
	\affiliation{Dipartimento di Fisica ``E. Pancini'', Universit\`{a} di Napoli ``Federico II'', Napoli, Italy.}
	
	
	\author{Salvatore Capozziello\orcidSalvatore\!}
	\email{capozziello@na.infn.it}
	\affiliation{Scuola Superiore Meridionale, Largo San Marcellino 10, I-80138, Napoli, Italy.}
	\affiliation{INFN Sez. di Napoli, Compl. Univ. di Monte S. Angelo, Edificio G, Via	Cinthia, I-80126, Napoli, Italy.}
	\affiliation{Dipartimento di Fisica ``E. Pancini'', Universit\`{a} di Napoli ``Federico II'', Napoli, Italy.}
	\affiliation{Laboratory of Theoretical Cosmology, Tomsk State University of Control Systems and Radioelectronics (TUSUR), 634050 Tomsk, Russia.}
	
	
	\author{Francisco S. N. Lobo\orcidFrancisco\,}
	\email{fslobo@fc.ul.pt}
	\affiliation{Instituto de Astrof\'{i}sica e Ci\^{e}ncias do Espaco, Faculdade de Ci\^encias da Universidade de Lisboa, Edif\'{i}cio C8, Campo Grande, PT-1749-016, Lisbon, Portugal.}
	\affiliation{Departamento de F\'{i}sica, Faculdade de Ci\^{e}ncias, Universidade de Lisboa, Edif\'{i}cio C8, Campo Grande, PT-1749-016 Lisbon, Portugal.}

\begin{abstract}
The Big Bang initial singularity problem can be solved by means of bouncing solutions. In the context of extended theories of gravity, we will look for covariant effective actions whose field equations contain up to fourth-order derivatives of the metric tensor. In finding such bouncing solutions, we will make use of an order reduction technique based on a perturbative approach. Reducing the order of the field equations to second-order, we are able to find solutions which are perturbatively close to General Relativity. We will build the covariant effective actions of the resulting order reduced theories.
\end{abstract}

\maketitle

\section{Introduction}
The classical theory of Big Bang describes the beginning of the Universe as an initial singularity wherein the temperature and the density assume an infinite value. A~way to overcome this unphysical situation is to invoke quantum gravity effects, which become dominant over the other fundamental interactions. An~attempt to solve the Big Bang singularity is the Loop Quantum Cosmology (LQC)\z~\cite{Ashtekar:2011ni,Bojowald:2012xy,Bojowald:2018sgf}. It reframes the initial singularity as a quantum big bounce generated by a short-range repulsive force which is dominant (and relevant) exclusively at Planck regime. Therefore, this implies the existence of a maximal value of the energy density at which the bounce occurs. 
As a result, the~bounce can be described by an effective Friedmann equation which reads as follows:
\begin{equation}\label{eq:lqc}
	H^2=\frac{1}{3}\kappa \rho\left(1-\frac{\rho}{\rho_c}\right)\,,
\end{equation}
where the LHS is the Hubble rate squared, where $H=\dot{a}/a$, $a=a(t)$ is the scale factor, and~the {\em dot} denotes a derivative with respect to the cosmological time $t$. In~the RHS, we have the constant $\kappa=8\pi\,G_{N}/c^4$, with~$G_N$ and $c$ being the Newton constant and the speed of light, respectively, the~energy density $\rho$, and~the maximum value of the energy density $\rho_c = c^2\sqrt{3}/(32\pi^2\gamma^3 G_N \ell^2_{P})$, with~$\gamma\approx 0.2375$ and $\ell_{P}=\sqrt{\hslash G_{N}/c^3}$ being respectively the Barbero--Immirzi parameter and the Planck length. Throughout this work, we are going to adopt the Planck units, where $c=\hslash=G_N=k_B=1$. 
The bounce occurs when energy density reaches its critical value, for~$\rho=\rho_c$, which correspond to $H^2=0$.

Notice that the quadratic term $\rho^2$ becomes significant just at energy densities comparable with the critical energy density. For~energies such that $\rho\ll\rho_c$, the~correction to the standard Friedmann equation $\propto\rho^2/\rho_c$ becomes negligible: this situation characterizes the energy range associated to the classical limit. The~quantum nature of this correction can be noticed approaching to the classical limit by making the limit $\hslash\rightarrow0$, which implies $\rho_c\rightarrow\infty$. Then, the~effective Friedmann equation ends up being the classical~one.	

Another condition for the bounce to take place is that $\ddot{a}/a>0$ at $\rho=\rho_c$, corresponding to an expansion phase after the contraction one\z~\cite{Singh:2006im,Taveras:2008ke}. As~a consequence, describing the matter contents a cosmological perfect fluid characterized by a constant barotropic equation of state for the pressure,  $p=w\rho$, obeying the standard conservation law $\dot{\rho}=-3H(1+w)\rho$ in a Friedmann--Lemaître--Robertson--Walker (FLRW) metric, it results that the barotropic constant must be $w>-1$. The~most interesting case is for $w=1$, which correspond to a massless scalar field in~LQC.

In this framework, we can write the second effective Friedmann equation as follows 
\begin{equation}\label{eq:lqc2}
	\frac{\ddot{a}}{a}=-\frac{1}{6}\kappa\,(\rho+3p)\left(1-2\,\frac{2\rho+3p}{\rho+3p}\,\frac{\rho}{\rho_c}\,\right)\,\,\longrightarrow\enspace \frac{\ddot{a}}{a}=-\frac{1}{6}\kappa\rho(1+3w)\!\left(1-2\,\frac{2+3w}{1+3w}\,\frac{\rho}{\rho_c}\,\right).
\end{equation}

A crucial characteristic of the Equation\z\eqref{eq:lqc} is that it is the standard Friedmann equation of General Relativity (GR) with an additive energy source. Therefore, it does not increase the GR degrees of freedom. In~this regard, it is possible to redefine the pressure and energy density in terms of new effective variables, given by
\begin{equation}
	\rho_{\rm{eff}}=\rho\left(1-\frac{\rho}{\rho_c}\right),\qquad p_{\rm{eff}}= p\left(1-2\frac{\rho}{\rho_c}\right)-\frac{\rho^2}{\rho_c}\,,
\end{equation}
and the Friedmann equations recover their standard form in terms of the new effective~variables.

Starting from the Equation\z\eqref{eq:lqc}, we can build up an effective action, in~the context of extended theories of gravity, such that it mimics the effective Friedmann equation of LQC but provides GR for $\rho_c\to\infty$.
However, this type of theories is characterized by higher-order derivatives of the metric tensor which, as~a rule, provide spurious degrees of freedom. A~way to face this inconvenience is to adopt an order reduction technique, which allows to obtain effective action perturbatively close to GR \z~\cite{Sotiriou:2008ya,Terrucha:2019jpm,Barros:2019pvc,Bajardi:2020fxh,Miranda:2021oig}.
The reduction technique is based on the redefinition of the Lagrangian as $\mathcal{L}=\sqrt{-g}(R+\epsilon\f)/2\kappa$, where $g$ is the determinant of the metric tensor, $R$ is the Ricci scalar, $\epsilon$ is a perturbative dimensionless parameter which indicates the deviation of the model from GR, and~$\f$ is a function of the curvature invariants. If~$\epsilon$ is set to zero, the~action recovers the standard Einstein--Hilbert one. Moreover, the~use of the order reduction technique is allowed at range of curvature such that $\epsilon\f \ll R$.

The field equations of an extended theory of gravity can be written as $G_{\mu\nu}=\kappa T_{\mu\nu}+\epsilon T^{(\f)}_{\mu\nu}$, where $T_{\mu\nu}$ is the energy-momentum tensor of matter, and $T^{( \f)}_{\mu\nu}$ is an additional effective energy-momentum tensor due to the presence of the function $\f$ inside the action (see Refs.\z~\cite{Capozziello:2011et, Nojiri:2010wj, Nojiri:2017ncd,Capozziello:2013vna,Capozziello:2014bqa,Capozziello:2018ddp}). Evaluating the $0$-$0$ component of field equations, we can derive the modified Friedmann equation of the theory. At~this point, we can connect the LQC correction of Equation\z\eqref{eq:lqc} to the additional contributions coming from the curvature energy-momentum tensor (characterizing extended theories of gravity) by requiring that the modified Friedmann equation mimics Equation\z\eqref{eq:lqc}.
This connection is done by using a perturbative approach, having GR as zeroth o. Once the perturbation parameter $\epsilon$ has been introduced, we compare the term $-\kappa\rho^2/3\rho_c$ of Equation\z\eqref{eq:lqc} to $\epsilon \bar{T}^{(\f)}_{00}$, where $\bar{T}^{(\f)}_{00}$ is the $0$-$0$ component of the effective energy-momentum tensor evaluated at the zeroth perturbative order in $\epsilon$, in~a spatially flat Friedmann--Lemaître--Robertson--Walker (FLRW) background. 
The comparison provides a differential equation for $\f$. In~this way, we are able to obtain functional forms of $\f$. 

In the past, other works (see Refs.\z~\cite{Sotiriou:2008ya,Terrucha:2019jpm,Barros:2019pvc}) adopted the current approach to obtain functional forms of $\f=\f(R)$, $\f=\f(G)$, and~$\f=\f(R,G)$, being $G=Q-4P+R^2$ the Gauss-Bonnet topological term, where $R=g^{\mu\nu}R_{\mu\nu}$ is the Ricci scalar, $P=R_{\mu\nu}R^{\mu\nu}$ is the Ricci tensor squared, and~$Q=R_{\mu\nu\rho\sigma}R^{\mu\nu\rho\sigma}$ is the Kretschmann scalar. 
An interesting generalization of these cases consists in taking into account $\f=\f(R,P,Q)$, (see Ref.\z~\cite{Miranda:2021oig}). In~this way, we can deal with a general fourth-order theory of gravity which is perturbatively close to~GR.

In Section\z\ref{method} we briefly describe the general approach based on the order reduction technique, highlighting its key steps, and~in Section\z\ref{general} we focus on the recent generalization of such an approach to $\f=\f(R,P,Q)$\z~\cite{Miranda:2021oig}. In~particular, we will focus on the case $w=1$ which corresponds to a massless scalar field in~LQC.

\section{Order Reduction Technique in FLRW~Background}\label{method}

Let us consider a general theory of gravity in a 4-dimensional spacetime of the form
\begin{equation}
	\label{action}
	\mathcal{S}= \frac{1}{2\kappa}\int d^4 x \sqrt{-g}\, (R+\epsilon\f)  + \mathcal{S}_m (g_{\mu\nu},\psi)\,,
\end{equation}
where $\f$ is a function of the curvature invariants (e.g., $\f=\f(R)$, $\f=\f(G)$, $\f=\f(R,G)$, $\f=\f(R,P,Q)$) and $\mathcal{S}_m (g_{\mu\nu},\psi)$ is the matter action, with~$\psi$ denoting the ensemble of the matter fields which are minimally coupled to the metric tensor. Then, varying the Equation\z\eqref{action} with respect to the metric tensor, we obtain the field equations
\begin{equation}\label{feq}
	G_{\mu\nu}=\kappa T_{\mu\nu}+\epsilon T^{( \f)}_{\mu\nu}\,,
\end{equation}
where, $T_{\mu\nu}$ is defined in the standard manner, namely,
\begin{equation}
	T_{\mu\nu}=-\frac{2}{\sqrt{-g}} \frac{\delta S_{m}}{\delta g^{\mu\nu}}\,,
\end{equation}
and the effective energy-momentum tensor $T^{(\f)}_{\mu\nu}$ is
\begin{equation}
	T^{(\f)}_{\mu\nu}=-\frac{1}{\sqrt{-g}} \frac{\delta }{\delta g^{\mu\nu}}\int d^4 x \sqrt{-g}\f\,.
\end{equation}

Once we set on which invariants $\f$ depends, then we obtain the explicit form of $T^{(\f)}_{\mu\nu}$ in terms of partial derivatives of $\f$ with respect to its variables, and~covariant derivative of the invariants and curvature~tensors. 

Extended theories of gravity such Equation\z\eqref{action} are characterized by the presence of further degrees of freedom. To~restore the original degrees of freedom of GR, we can consider the GR field equations as the zeroth perturbative order of our theory (i.e., $\epsilon=0$) and use them to express the curvature tensors in terms of the matter contents. We indicate the zeroth order with a \textit{bar}, and~we refer to the barred quantities as their \av reduced\cv~form. 

The Equation\z\eqref{action} at the zeroth-order reduces to Einstein field equations:
\begin{equation}
	\bar{R}_{\mu\nu}-\frac{1}{2}g_{\mu\nu}\bar{R}=\kappa T_{\mu\nu}\,\quad\Rightarrow\quad\bar{R}_{\mu\nu}=\kappa T_{\mu\nu}+\frac{1}{2}g_{\mu\nu}\bar{R}\,. 
\end{equation}

Then, taking the trace of the above equations we obtain the expressions of the reduced Ricci scalar and the reduced Ricci tensor in terms of the matter contents:
\begin{align}
	&\bar{R}=-\kappa T\,,\label{scalar}\\
	&\bar{R}_{\mu\nu}=\kappa T_{\mu\nu}-\frac{\kappa}{2}g_{\mu\nu}T\,,\label{ricc}
\end{align}
where $T=g^{\mu\nu}T_{\mu\nu}$ is the trace of the energy-momentum~tensor.

In particular, we deal with a spatially flat FLRW background 
\begin{equation}
	ds^2 = -dt^2 +a(t)^2 \left[ dr^2 +r^2\left( d\theta^2 + \sin^2\theta\, d\phi^2 \right) \right]\,,
\end{equation}
which has the characteristic to be conformally flat. Therefore, the~Weyl tensor is identically null, and~we can express the Riemann tensor in terms of the Ricci tensor and the Ricci scalar. Then, the~reduced form of the Riemann tensor turns out to be
\begin{equation}
	\bar{R}_{\mu\nu\alpha\beta} =-\frac{\kappa}{2}\left( g_{\mu\beta}T_{\alpha\nu}+g_{\nu\alpha}T_{\beta\mu} - g_{\mu\alpha}T_{\beta\nu}-g_{\nu\beta}T_{\alpha\mu} \right) -\frac{\kappa}{3} T\left( g_{\mu\alpha}g_{\nu\beta}-g_{\mu\beta}g_{\alpha\nu}\right) \,. \label{riem}
\end{equation}

Finally, we consider an energy-momentum tensor associated to a homogeneous isotropic cosmological perfect fluid, i.e.,~$T_{\mu\nu} = \left( \rho + p \right)u_{\mu}u_{\nu} + p\,g_{\mu\nu}$, where the 4-velocity of the fluid $u^{\mu}$ is normalized as $u_{\mu} u^{\mu}=-1$, while $p$ and $\rho$ are the isotropic pressure and energy density, respectively. Here, we assume a barotropic equation of state $p = w\rho$, with~a  barotropic constant $w$. Thus, it results that $T_{\mu\nu} = \rho\left( 1 + w \right)u_{\mu}u_{\nu} + w\rho\,g_{\mu\nu}\,$. Then, the~energy-momentum conservation law, $\nabla_{\mu}T^{\mu\nu}=0$, provides the  continuity equation $\dot{\rho}= -3H \left( 1+w \right)\rho\,.$

Now, we have all the tools to rewrite the modified Friendann equation, i.e.,~the $0$-$0$ component of the Equation\z\eqref{feq}, in~its reduced form\,,
\begin{equation}\label{mfriedmann}
	H^2(t)=\frac{1}{3}\kappa\rho+\frac{1}{3}\epsilon T^{(\f)}_{00}\quad\longrightarrow\quad H^2(t)=\frac{1}{3}\kappa\rho+\frac{1}{3}\epsilon \bar{T}^{(\f)}_{00}\,,
\end{equation}
and comparing Equation\z\eqref{eq:lqc} with the Equation\z\eqref{mfriedmann}, we obtain the following equation
\begin{equation}\label{mlqc}
	\bar{T}^{(\f)}_{00}=-\frac{\kappa\rho^2}{\epsilon\rho_c}\,.
\end{equation}

The above equation represents a differential equation for $\bar{\f}$ once we specify the variables on which $\f$ depends. However, the~Equation\z\eqref{mlqc} is not solvable in general. Of~course, the~solvability depends on the variables of $\f$, and~on the relation between $\rho$ and the reduced variables. Moreover, it is possible to find a general integral only if $\f$ depends on just one variable (i.e., $\f=\f(R)$, $\f=\f(G)$), otherwise it is necessary to make some ansatz on the possible functional forms of $\f$.

Let us implement this method with the simplest extension of GR: $f(R)\rightarrow R+\epsilon\f(R)$. In~this case, we have:
\begin{equation}
	\begin{array}{ll}
		&T^{(\f)}_{\mu\nu}=\dfrac{1}{2}\f g_{\mu\nu}-\f'R_{\mu\nu}+[\nabla_\mu\nabla_\nu-g_{\mu\nu}\Box]\f'\quad\Rightarrow\vspace{3pt}\\
		&\bar{T}^{(\f)}_{00}=-\dfrac{1}{2}\bar{\f}-\dfrac{1}{2}(1+3w)\kappa\rho\bar{\f}'+3(1+w)(1-3w)\kappa^2\rho^2\bar{\f}''\,,
	\end{array}
\end{equation}
where $\nabla_\mu$ indicates the Levi--Civita covariant derivative, box is defined as $\Box=\nabla_\mu\nabla^\mu$, and~the prime stands for the derivative with respect to $\bar{R}$, with~$\bar{\f}=\f(\bar{R})$.

Then, using the Equation\z\eqref{scalar} to write $\kappa\rho=R/(1-3w)$, and~substituting it inside the above equation, we get the following differential equation for $\f=\f(R)$:
\begin{equation}
	(1-3w)\f(R)+(1+3w)R\f'(R)-6(1+w)R^2\f''(R)=\frac{2R^2}{(1-3w)\epsilon\kappa\rho_c}\,.
\end{equation}

The latter is a second order non-homogeneous Euler–Cauchy equation.Once we fix the barotropic constant $w=1$, we obtain the following solution
\begin{equation}
	\f(R)=\frac{R^2}{18\epsilon\kappa\rho_c}+A\,R^{\frac{1}{6}\left(4+\sqrt{10}\right)}+B\,R^{\frac{1}{6}\left(4-\sqrt{10}\right)}\,,
\end{equation}
where $A$ and $B$ are integration constants.

The procedure is the same for $R+\epsilon\f(G)$, where $\f$ depend on just one variable. Otherwise, in~the case of more than one variable, e.g.,~$f(R,G)\rightarrow R+\epsilon\f(R,G)$, we are obliged to make some ansatz on $\f$ to find explicit solutions of Equation\z\eqref{mlqc}. Thus, to~better understand what happens when $\f$ is a multivariable function, we apply the method to a general fourth-order theory of gravity, which represents the generalization of all the works present in literature, until~today.

\section{Loop-Inspired Fourth-Order Theory of~Gravity}\label{general}

Here, we want to apply the order reduction technique to general fourth-order theory of gravity having the following form 
\begin{equation}
	f(R,P,Q)\rightarrow R+\epsilon\f(R,P,Q)\,.
\end{equation}

In order to obtain the reduced form of the invariants $P$ and $Q$, we use the Equations\z\eqref{scalar},\z\eqref{ricc} and\z\eqref{riem}. Thus, it results
\begin{align}\label{PQ}
	\bar{P}=\kappa^{2}T^{\mu\nu}T_{\mu\nu}\,,\qquad\bar{Q}=-\frac{1}{3}\kappa^2 T^2+2\kappa^2 T_{\mu\nu}T^{\mu\nu}\,,
\end{align}
where, $T_{\mu\nu} = \rho\left( 1 + w \right)u_{\mu}u_{\nu} + w\rho\,g_{\mu\nu}\,$ in a FLRW~spacetime. 

In this case, the~reduced form of the effective energy-momentum tensor is
\begin{equation}
	\begin{array}{ll}
			&\bar{T}^{(\f)}_{\mu\nu}=\dfrac{1}{2}g_{\mu\nu}\bar{\f}-\bar{R}_{\mu\nu}\bar{\f}_{,\bar{R}}-\nabla_{\mu}\nabla_{\nu}{\bar{\f}}_{,\bar{R}}+g_{\mu\nu}\Box {\bar{\f}}_{,\bar{R}}-2\bar{\f}_{,\bar{Q}}\bar{R}_{\alpha\beta\gamma\mu}\bar{R}^{\alpha\beta\gamma}{}_{\nu}-2{\bar{\f}}_{,\bar{P}}\bar{R}^{\alpha}{}_{\mu}\bar{R}_{\alpha\nu}\vspace{3
				pt}\\
			&+4\nabla_{\alpha}\nabla_{\beta}(\bar{\f}_{,\bar{Q}}\bar{R}^{\alpha}{}_{(\mu\nu)}{}^{\beta})+2\nabla_{\alpha}\nabla_{\beta}(\bar{\f}_{,\bar{P}}\bar{R}^{\alpha}{}_{(\mu}\delta^{\beta}{}_{\nu)})-\Box(\bar{\f}_{,\bar{P}}\bar{R}_{\mu\nu})-g_{\mu\nu}\nabla_{\alpha}\nabla_{\beta}(\bar{\f}_{,\bar{P}}\bar{R}^{\alpha\beta})\,,
	\end{array}
\end{equation}
where $\bar{\f}=\f(\bar{R},\bar{P},\bar{Q})$, the~comma denotes a partial derivative, i.e.,~$\bar{\f}_{,\bar{R}}=\partial{\f}/\partial{\bar{R}}$, while the subscripted parenthesis indicate the symmetric part of a tensor with respect to the indices inside of them, i.e.,~$S_{(ab)c}=(S_{abc}+S_{bac})/2$. Then, the~Equation\z\eqref{mlqc} has the following explicit form:
\begin{equation}\label{de}
		\begin{array}{ll}
			&-\dfrac{1}{6}\bar{\f}+\dfrac{1}{6}  (-3 w-1) \kappa\rho\, \bar{\f}_{,\bar{R}}+\dfrac{1}{3}  \left(-3 w^2-3 w+2\right) \kappa^2\rho^2 \bar{\f}_{,\bar{P}}\vspace{3pt}\\
			&+\dfrac{1}{9}  \left(-9 w^2-6 w+7\right) \kappa^2\rho^2\, \bar{\f}_{,\bar{Q}}- (w+1) (3 w-1) \kappa^2\rho^2\, \bar{\f}_{,\bar{R},\bar{R}}\vspace{3pt}\\
			&-4 w (w+1) \left(3 w^2+1\right)  \kappa ^4\rho^4\, \bar{\f}_{,\bar{P},\bar{P}}-\dfrac{4}{9} \left(27 w^4+54 w^3+48 w^2+26 w+5\right) \kappa^4 \rho^4\,\bar{\f}_{,\bar{Q},\bar{Q}}\vspace{3pt}\\
			&+2  (w+1) \left(6 w^2-w+1\right) \kappa^3\rho^3\,\bar{\f}_{,\bar{\bar{R}},\bar{P}}+\dfrac{4}{3}  (w+1) \left(9 w^2+3 w+2\right) \kappa^3\rho^3\,\bar{\f}_{,\bar{R},\bar{Q}}\vspace{3pt}\\
			&-\dfrac{4}{3}  (w+1) \left(18 w^3+9 w^2+8 w+1\right) \kappa^4\rho^4\,\bar{\f}_{,\bar{P},\bar{Q}}=-\dfrac{\kappa^2\rho^2}{3\epsilon\kappa\rho_c}\,.
	\end{array}
\end{equation}

The Equation\z\eqref{de} is a non-linear second-order partial differential equations for a three-variable function. Moreover, there is not a unique way to write $\kappa\rho$ in terms of $\bar{R}$, $\bar{P}$, and~$\bar{Q}$. Therefore, it is not possible to determine the general form of $\bar{\f}$.
Thus, we need to propose some specific \textit{ansatz} on $\f$ and investigate the conditions under which $\f$ can mimic a bouncing~universe.

\subsection{Solution 1: Definition of a Single~Variable}
A way to obtain functional forms of $\f$, it is to transform Equation\z\eqref{de} in an single variable differential equation, by~defining a new single variable that depends on $R$, $P$ and $Q$: $X=X(R,P,Q)$. 

In this regard, let us define the simplest, but~still general, dimensionless variable of $R$, $P$, and~$Q$:
\begin{equation}
	f(R,P,Q)= R+\epsilon\f(X),\qquad{\rm with} \qquad \enspace X=c_1|R|^{2\alpha}+c_2P^\alpha+c_3Q^\alpha\label{ansatz1}
\end{equation}
where $\alpha$ is a dimensionless real parameter, and~the constants $\{c_i\}$ with $i=1,2,3$ have dimension of $[c_i]=[\rho]^{-2\alpha}$ such that $X$ is~dimensionless.

Now, we can use the assumption\z\eqref{ansatz1} to solve Equation\z\eqref{de}, by~writing the reduced form of $X$, which is given by
\begin{equation}
	\bar{X}= c\, (\kappa  \rho )^{2 \alpha }\,,\quad{\rm where}\quad c= c_1 \left| 1-3 w  \right| ^{2 \alpha }+c_2 \left(3 w  ^2+1\right)^{\alpha }+c_3 \,3^{-\alpha }\left(9+6w+5\right)^{\alpha }\,.\label{x}
\end{equation}

Notice, we are assuming to work with all possible configurations of parameters $\{w>-1,\alpha,c_i\}$ corresponding to a non-vanishing $\bar{X}$, namely, $c\neq0$ in Equation\z\eqref{x}.

It is straightforward to see that Equation\z\eqref{de} gives the following differential equation for $\bar{\f}=\f(|\bar{X}|)$:
\begin{equation}\label{dex}
	-\frac{1}{6} \bar{\f}+\frac{a}{c}\,|\bar{X}| \bar{\f}' + \frac{b}{c^2}\, |\bar{X}|^2 \bar{\f}''=-\frac{\left|\bar{X}\right|^{1/\alpha}}{3\epsilon\kappa\rho_c|c|^{1/\alpha}}
\end{equation}
where the prime denotes the derivative with respect to $\bar{X}$, and~$a$ and $b$ are coefficients depending on $w$, $\alpha$, and~$\{c_i\}$. 

The Equation\z\eqref{dex} is a second order non-homogeneous Euler--Cauchy equation. For~this reason, we already know all possible solutions, without~entering too much into values of the parameters $\{w,\,\alpha,\,c_i\}$. Notably, the~ensemble of the possible particular solutions, depending on $\{w,\,\alpha,\,c_i\}$ can be written in a single expression, as~follows
\begin{equation}
	\f(X)\propto |X|^{1/\alpha}\left(\ln{|X|}\right)^\beta\,,
\end{equation}
where $\beta=0,1,2$ is null only if $1/\alpha$ is not a root of the characteristic equation associated to Equation\z\eqref{dex}, otherwise $\beta$ is the multiplicity of the~root.

In particular, we fix $\alpha=1$ for simplicity, and, henceforward, we also set $w=1$. Thus, the~space of solutions is considerably reduced, and~we can summarize the solutions as follows:
\begin{itemize}
	\item  For $6c_1+3c_2+4c_3\neq0$ and $3 c_1+c_2+c_3\neq0$, we obtain
	\begin{equation}\label{complete}
		\f({X})=\frac{\left|{X}\right|}{6 \left(3 c_1+c_2+c_3\right)  \epsilon \kappa \rho _c}+A\,\left|{X}\right|^{(-b-\sqrt{\Delta})/(2a)}+B\,\left|{X}\right|^{(-b+\sqrt{\Delta})/(2a)}\,,
	\end{equation}
	where
	\begin{equation}\label{parameters}\begin{array}{cc}
			&a=6\left(6 c_1+3 c_2+4 c_3\right),\qquad b=-24 c_1-15 c_2-22 c_3,\vspace{3pt}\\
			&\Delta=360 c_1^2+396 c_2 c_1+552 c_3 c_1+117 c_2^2+244 c_3^2+336 c_2 c_3\,,
	\end{array}\end{equation}
	with $A$ and $B$ being integration constants.
	\item For $6c_1+3c_2+4c_3\neq0$ and $3 c_1+c_2+c_3=0$ we obtain
	\begin{equation}
		\f({X})= \frac{3 \,|X|\ln{|X|}}{20\left(6 c_1+c_2\right) \epsilon \kappa \rho _c}+A\, |X|^{1/6}+B\,|X|,
	\end{equation}
	where $A$ and $B$ are integration constants.
	\item For $6c_1+3c_2+4c_3=0$ and $c_2+2 c_3\neq 0$, we obtain
	\begin{equation}
		\f({X})=-\frac{|X|}{3 \left(c_2+2 c_3\right) \epsilon \kappa \rho _c}+A\,|X|^{1/4}\,,
	\end{equation}
	where $A$ is an integration constant.
\end{itemize}\vspace{10pt}

In the next two subsections, we propose two particular forms for $\f(R,P,Q)$ inspired by the solutions obtained in the single variable case: a power-law form and a logarithmic form, respectively.

\subsection{Solution 2: Power-Law~Form}
Let us assume $\f$ has the following power-law:
\begin{equation}
	\f(R,P,Q)=c_0\, |R|^{\alpha}\,P^{\beta}\,Q^{\gamma}\,|c_1R^2+c_2P+c_3Q|^{\delta}\label{ansatz2}\,,
\end{equation}
where $\{c_i\}$, with~$i=1,2,3$, and~$\alpha,\,\beta,\,\gamma,\,\delta$ are real dimensionless parameters, while $c_0$ has dimension $[c_0]=[\rho]^{1-(\alpha+2\beta+2\gamma+2\delta)}$, such that $\f$ has the dimension of an energy~density. 

Our aim is to express the constant $c_0$ as a combination of the other parameters, $\{c_i,\,\alpha,\,\beta,\,\gamma,\,\delta\}$, such that Equation\z\eqref{ansatz2} is a solution of the Equation\z\eqref{de}. 

Here, we are excluding all the cases in which the reduced form of Equation\z\eqref{ansatz2} is identically zero, which constrain the space of the possible values of the parameters of the~solution.

Substituting Equation\z\eqref{ansatz2} in Equation\z\eqref{de}, we obtain what we call dimensional constraint:
\begin{equation}
	\alpha =2- 2(\beta +\gamma +\delta)\,.
\end{equation}

Using the above constraint, it is possible to obtain a formula for $c_0$ in terms of the other parameters. Moreover, one can verify that, also removing the dependence on $R$ from Equation\z\eqref{ansatz2}, a~power-law form of $\f$ is not a solution for the Equation\z\eqref{de} if $w=1/3$, which corresponds to $\bar{R}=0$.

In particular, for~the case of interest $w=1$, the~constant $c_0$ takes the form:
\begin{equation}
	c_0= -\frac{ 3^{\gamma +\delta -1}\,5^{1-\gamma } \left(3 c_1+3 c_2+5 c_3\right) \left| 3 c_1+3 c_2+5 c_3\right| {}^{-\delta }}{2 \epsilon \kappa \rho _c \left(\left(3 c_1+3 c_2+5 c_3\right) (10 \beta +12 \gamma -15)+30 \left(c_2+2 c_3\right) \delta\right)}\,.
\end{equation}
%

\subsection{Solution 3: Logarithmic~Dependence}
Let us assume $\f$ has the following logarithmic dependence:
	\begin{equation}
		\f(R,P,Q)=c_0\, |R|{}^{\alpha_1}\,P{}^{\beta_1}\,Q{}^{\gamma_1}\,|c_1R^2+c_2P+c_3Q|{}^{\delta_1}\ln{\left(c_{\rho}\,|R|{}^{\alpha_2}\,{P}{}^{\beta_2}\,{Q}{}^{\gamma_2}\,|c_4{R}^2+c_5{P}+c_6{Q}|{}^{\delta_2}\right)}\label{ansatz3}
	\end{equation}
where $\{c_i,\,\alpha_j,\,\beta_j,\,\gamma_j,\,\delta_j\}$, with~$i=1,2,3$ and $j=1,2$, are real dimensionless parameters, while the constant $c_{\rho}$ has dimension $[c_{\rho}]=[\rho]^{-\alpha_2-2(\beta_2+\gamma_2+\delta_2)}$, and~it is used to render dimensionless the argument of the~logarithm.

It is necessary to reiterate that we are excluding all these cases in which the above function is identically zero, which constrain the space of the possible values of the parameters of the~solution.

Replacing Equation\z\eqref{ansatz3} in Equation\z\eqref{de}, we obtain a dimensional constraint $\alpha_1 =2- 2(\beta_1 +\gamma_1 +\delta_1)$. Moreover, for~Equation\z\eqref{ansatz3} to be a solution of Equation\z\eqref{de}, we must impose the following condition:
	\begin{equation}\begin{array}{c}\label{log}
			\left[c_2 \left(9 w ^2+3\right)+c_3 \left(9 w ^2+6 w +5\right)+3 c_1 (1-3 w )^2\right]\times\vspace{3pt}\\
			\times\left[-2 \beta _1 (3 w +1) \left(9 w ^2+6 w +5\right)-12 \gamma _1 (3 w +1) \left(3 w ^2+1\right)+3 \left(3 w ^2+1\right) \left(9 w ^2+6 w +5\right)\right]\vspace{3pt}\\
			-6 \delta _1\left(c_2+2 c_3\right)  (3 w +1) \left(3 w ^2+1\right) \left(9 w ^2+6 w +5\right)=0\,.
	\end{array}\end{equation}

The Equation\z\eqref{log} guarantees that $c_0$ is a constant. When it holds, it is possible to express $c_0$ in terms of the other parameters such that the ansatz\z\eqref{ansatz3} mimics the bouncing universe of LQC.
In particular, fixing $w=1$, we obtain
\begin{equation}
	c_0= \frac{5^{1-\gamma_1} \left(3 c_4+3 c_5+5 c_6\right) 3^{\gamma_1+\delta_1} \left| 3 c_1+3 c_2+5 c_3\right| {}^{-\delta_1}}{2 \epsilon \kappa \rho_c \left(\left(3 c_4+3 c_5+5 c_6\right) \left(35 \alpha_2+40 \beta_2+34 \gamma_2\right)+10 \left(21 c_4+12 c_5+17 c_6\right) \delta_2\right)}\,.
\end{equation}
%

\section{Discussion and~Conclusions}
In the framework of $f(R,P,Q)$ extended theory of gravity, we provide specific cosmological model reproducing the effective Friedmann equation of~LQC. 

The main tool is an order reduction technique, showed in the Section\z\ref{method}. The~first step is to introduce a perturbative parameter $\epsilon$, rewriting the theory as $f(R,P,Q)\rightarrow R+\epsilon\f(R,P,Q)$. The~second step is  to express the geometric variables in terms of matter content, i.e.,~at the zeroth perturbation order. In~this way, we are able to find additive contributions to the Ricci scalar building a theory which is perturbatively close to GR. These terms can be interpreted as corrections to GR at the energy scales close to the Planck scale. This interpretation is consistent with GR prediction at low-energy scales where these terms are completely negligible. However, because~the semi-classical nature of our approach, our treatment is valid only in proximity of the bounce, i.e.,~$\rho\sim 0.1\rho_c$, but~does not in correspondence of it.
Therefore, we find building blocks for an effective action which can mimic the effective Friedmann equation of LQC far enough from the bounce.
However, approaching too close to the bounce (i.e., $R\sim\epsilon\f$), the~perturbative approach is no more valid. Nevertheless, one could invoke loop quantum gravity solutions, and~a final fundamental theory (once discovered) would give the features of the bouncing solution described by our action and described by loop quantum gravity at the~bounce.

Our analysis generalizes the results obtained in previous works for $f(R)$, $R+f({\cal G})$ and $f(R,{\cal G})$ theories of gravity\z~\cite{Sotiriou:2008ya,Terrucha:2019jpm,Barros:2019pvc}, providing the same results in appropriate limits (see Ref.\z~\cite{Miranda:2021oig} for more details). In such a context, we have found bouncing solutions assuming specific ansatz and obtaining algebraic constraints on the free parameters of the models. 
	However, since we are just dealing with the theoretical framework leading to a bounce by using the reduction approach, we are not taking into account any physical constraint coming from observations.

Further generalizations can be obtained by considering other extended theories of gravity, and~the context of \av modified LQC Friedmann equations\cv\!\!\z~\cite{Li:2019ipm,Ribeiro:2021gds}. Therefore, in~the future we also aim to extend the present analysis to the general context of modified~LQC. 

To conclude, it is worth to say that we focus our attention just on cosmological bouncing models. However, there are a lot of alternative models to solve the initial singularity. Among~these models, the~\av emergent universe\cv deserves to be mentioned\z~\cite{Ellis:2002we,Ellis:2003qz,Guendelman:2014bva}. Here, scalar fields cure the initial singularity, evolving then into viable inflation.

\acknowledgments{M.M., D.V. and~S.C. acknowledge support by the {\it Istituto Nazionale di Fisica~Nucleare} (INFN) ({\it iniziative specifiche} MOONLIGHT-2,  TEONGRAV, and~QGSKY). F.S.N.L. acknowledges support from the Fundac\~{a}o para a Ci\^{e}ncia e a Tecnologia (FCT) Scientific Employment Stimulus contract with reference CEECINST/00032/2018, and~funding from the research grants No.~UID/FIS/04434/2020, No. PTDC/FIS-OUT/29048/2017 and No. CERN/FIS-PAR/0037/2019. }

\end{document}